# A Precise Program Phase Identification Method Based on Frequency Domain Analysis

Hsuan-Yi Lin
National Tsing Hua University
Hsinchu, Taiwan
mark126688@hotmail.com

Ren-Song Tsay
National Tsing Hua University
Hsinchu, Taiwan
rstsay@cs.nthu.edu.tw

*Abstract*—**In this paper, we present a systematic approach that transforms the program execution trace into frequency domain and precisely identifies program phases. The analyzed results can be embedded into program code to mark the starting point and execution characteristics, such as CPI (Cycles per Instruction), of each phase. The so generated information can be applied to runtime program phase prediction. With the precise program phase information, more intelligent software and system optimization techniques can be further explored and developed.**

## 1. Introduction

As designers demand better system optimization results, a consensus is that target application runtime behaviors should be considered in the optimization process. Particularly for embedded system designs, which by definition are dedicated for specific applications, an accurate application behavior model is known to be critical for performance optimization.

In terms of behavior pattern, we observe that in practice, most applications have repetitive execution phases which alternatively occur in each execution run. In general, each execution phase exhibits consistent performance value. In Figure 1 we show an illustrative example which has three different program phases, as indicated by labels A, B, and C, which alternatively occur through the execution run and each phase clearly performs at a certain CPI (Cycles per Instruction) value. In practice, each different program phase implies specific data computing and access behaviors.

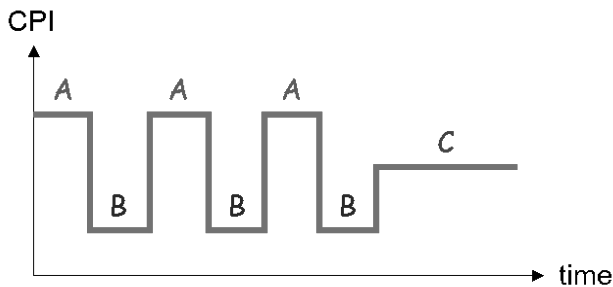

Figure 1. An example with three program phases. Each phase has a different CPI value.

Many researchers have developed various program phase prediction techniques in attempt to consider and take advantage of the repetitive program behaviors for system performance and power optimization [1, 2, 3, 4, 5, 6]. Although encouraging results were observed, there was no clear agreement on defining what a program phase is.

These past approaches in general attempted to divide the application execution trace into fixed-length execution segments, and merge those consecutive segments of *same* performance metric into a phase. The *fixed length* can be a time quantum in terms of a fixed number of instructions, or a fixed number of basic blocks executed. The advantage of having fxied-length segments is that the evaluation metric and final analysis can be conveniently performed. Nevertheless, a major issue is that the fixed-length segments may not match well with the actual phase boundries, and hence phases of shorter length are ignored and phases with boundaries in between time quantums can be misjudged. Another dielama is that before performing analysis no useful information can be provided for *optimal* fixed length decision. In fact, we will see later generally there is no single optimal time quantum length for all cases.

Therefore, our goal is to try to capture the basic program phase building blocks and then develop a systematic approach for program phase identification with no ambiguity.

Most importantly, we observe a fact that the repetitive behaviors can be easily identified in frequency domain analysis. Essentially, we borrow the idea of spectrum analysis which is regularly used in signal processing field. We find that usually the low frequency spectrum implies the duration of the main program phase. Accordingly, we develop a unique and effective frequency-domain analysis technique for precise program phase identification.

Furthermore, according to our observation in practice, generally the application behaviors follow closely with the code functions executed. Thus, program phase is highly correlated to program code structure. With this hint, we propose an idea that perform program phase analysis process mainly on basic blocks but with the hints of loops and functions to precisely identify phase boundaries.

A basic block is essentially a straight-line code sequence with no branches except to the entry and from the exit. This restricted form makes a basic block highly amenable to analysis. Loop and function are another common code structures, which usually exhibit stable behavior in repeated executions.

Basically, we try to leverage the fact that basic blocks are the most basic execution units usually with consistent behavior within a program. Therefore, we first compute the average CPI value of each basic block and form the performance trace using basic blocks as time index. In this way, we conveniently resolve the granularity issue often associated with the time-quantum approaches. More precise calculation process is to be elaborated later in this paper.

The contributions of the proposed approach are as the following. First, the proposed frequency-domain analysis method allows precisely identification of program phases, including specific phase starting point, execution length, and behavior characteristics. Additionally, using basic blocks for analysis, we clarify the association between program phases and loop/function code structures. Our model greatly improves the prediction accuracy and provides insights for optimization. Finally, our analyzed results can be embedded into the target application code used for highly desired runtime optimization purpose.

The rest of this paper is organized as follows. In Section 2, we review and discuss related work. In Section 3, we present a systematic method for precise program phase identification through frequency domain analysis. In Section 4, we show experimental results to demonstrate the accuracy of our method and discuss the special cases. Finally, we give a brief conclusion in Section 5.

## 2. Related Work

Generally, there are two types of phase identification approaches. The first is a time-quantum-based approach which divides a program execution into fixed-length intervals, with each interval represents a phase segment which is to be merged into a phase. The second one is a program-structure-based approach, which uses the basic structures such as loops or functions to determine program phases. In general, the two aforementioned methods all exhibit certain defects, which we will discuss later.

### A. Time-quantum-based Approaches

A time quantum is a fixed-length contiguous execution interval and it is used to divide a program execution trace into non-overlapping quanta for phase analysis. Usually either a fixed number of instructions or basic blocks form an interval for evaluation. Then, quanta of *similar* performance metric, such as CPI, cache miss rate, branch miss rate, are grouped into one phase. Nevertheless, the term "*similar*" is not precisely defined and it implies certain error margin is allowed. The Basic Block Vector (BBV) approach [8, 9, 10] is one of the most representative time-quantum-based techniques.

The BBV method is motivated by the observation that the program phase behavior is highly correlated with the patterns of basic blocks. Mainly, the program behavior is a result of executing program code, which is composed of basic blocks. Therefore, BBV records the footprint of basic blocks in the execution trace and checks if the compositions (or named *signatures*) of any consecutive basic block vectors (a fixed number of basic blocks) display difference more than a given threshold value to identify the separation of two phases.

A common issue of this approach is that quanta often run across program phase boundaries and cause the so called transition phase issue [11]. Usually the quantum runs across an actual phase boundary is deemed as a unique phase of different performance measure and compared to that of the phase before and the phase after the boundary point.

Fang. et. al. [12] observed that phases should be of a hierarchical structure and proposed a Multi-Level Phase Analysis method that classified phases into fine-grain (an inner-loop) and coarse-grain (an outer-loop or function) phases. Basically, a coarse-grain phase is consisted of stably distributed fine-grain phases. The idea of having multi-level phases matches better with practical cases. However, with the differentiation of coarse-grain and fine-grain phases still cannot fully capture the general cases. In contrast, our approach provides a more systematic method to identify the full phase hierarchy.

One challenging issue of the fixed time-quantum-based approaches is that the fixed quantum length may not match with the program behavior rhythm, which can be irregular. In other words, the time-quantum-based method may not precisely capture actual program phases. Additionally, the fixed length constraint loses insights of the dynamic nature of real program phases. We hence propose in this paper an effective approach based on identifying the natural rhythm of the actual program behavior.

### B. Program-structure-based Approaches

Another type of approaches focuses on identifying phases based on loops and functions, which are commonly used structures in programs. Loops and functions normally are repeatedly executed with consistent performance behavior, which matches well with the desired characteristic of program phases. This method assumes that loops and functions are basic components of a phase, and the purpose is to find the transition that connects phases.

For instance, Huang. et. al. [1] followed the basic idea of phases and attempted to configure combinations of function calls (or subroutines) for performance or energy consumption improvement.

Lau. et. al. [13] used a Hierarchical Call-Loop Graph analysis technique to identify program phases from loops or functions (or named procedures). On the graph, each node represents a loop or a function. Each edge represents an execution path from a node to another node and is associated with a call count, the average and standard deviation of instruction count across various invocations. The edges with lower deviation numbers represent code segments of relative consistent performance behaviors and are each marked as a phase.

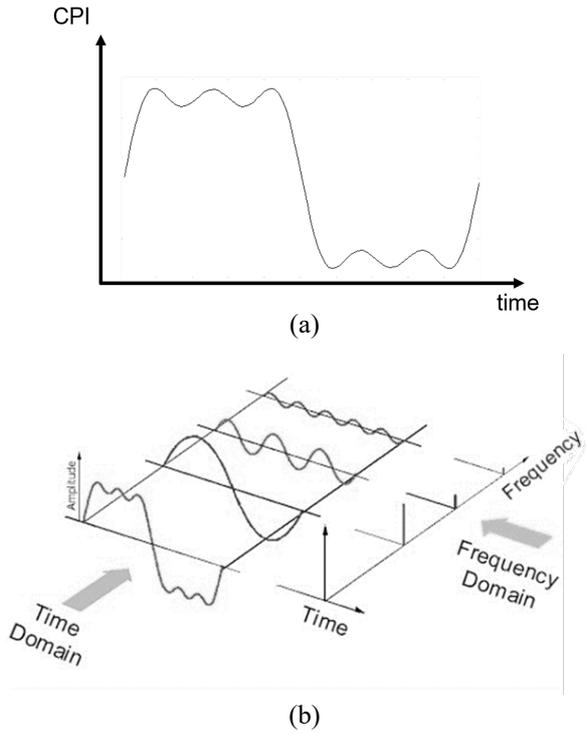

Figure 2. (a) A time-domain performance waveform example. (b) The low frequency spectrum in frequency domain corresponds to the major phase in time domain.

In contrast, Jiang. et. al. [14] did not explicitly identify program phases but attempted to identify the correlations among repeatedly executed loops and functions under various inputs. For instance, the analysis may find that a certain function's call count is almost always of a fixed ratio of a loop's trip-count under different inputs tested, then this statistical ratio is then applied to runtime program behavior prediction.

A common challenge to these program-structure-based approaches is that it is difficult to calculate a precise performance value (e.g. CPI) due to the variations of different execution runs.

In summary, the issue of above-mentioned existing approaches is that they all try to use a pre-conceived phase pattern, either a fixed phase length, a fixed basic block vector size, or connections to loops or functions etc. to determine a program phase. In fact, the pattern guessed generally may not match with the natural program phase patterns. Conversely, our idea is to identify program phases based on the inherent characteristic of basic blocks with the hints of program code structure.

Next, we will present our proposed frequency-domain-based approach.

## 3. Frequency-Domain Program Phase Analysis Method

Normally, human being can easily recognize repeated patterns, or program phases, simply by viewing the waveform. For example, the waveform in Figure 2(a) clearly has a high performance phase and a low performance phase. However, the challenge is to identify such patterns automatically through a systematic algorithm. To address this challenge, we leverage the fact that repeated patterns in time domain often show up as a particular frequency response. Therefore, we propose in the following a unique frequency-domain analysis approach for phase detection.

Our proposed approach is a practical method that can precisely identify phase starting point, phase length, and the average program performance of each phase. The approach also includes program loop and function structures for accurate phase identification. Now, we explain how the proposed idea works.

Note that since CPI (cycle per instruction) is a commonly used performance measure, we simply use CPI for program behavior characterization in the following discussion, but the approach is not limited from other performance measures.

### A. Frequency Domain Analysis

With a given program execution trace for analysis, we first create a time-domain waveform as shown in Figure 2(a), in which the vertical axis represents the performance value and the horizontal axis represents the time quantum sequence number. Each sample point on the waveform represents the performance value obtained at the corresponding instruction count index of the program execution.

Since a phase by definition is a frequently occurred pattern, particularly the major phase normally dominates the performance waveform. Once the time-domain performance waveform is transformed into frequency domain, the major phase normally shows up in low frequency spectrum, as illustrated in Figure 2(b). If fact, each frequency response corresponds to certain repeated program behavior patterns shown in the time-domain performance waveform.

In practice, a Fourier Transform tool [18] is used to perform frequency domain analysis (FDA). We use the example in Figure 3 to illustrate the proposed phase identification algorithm.

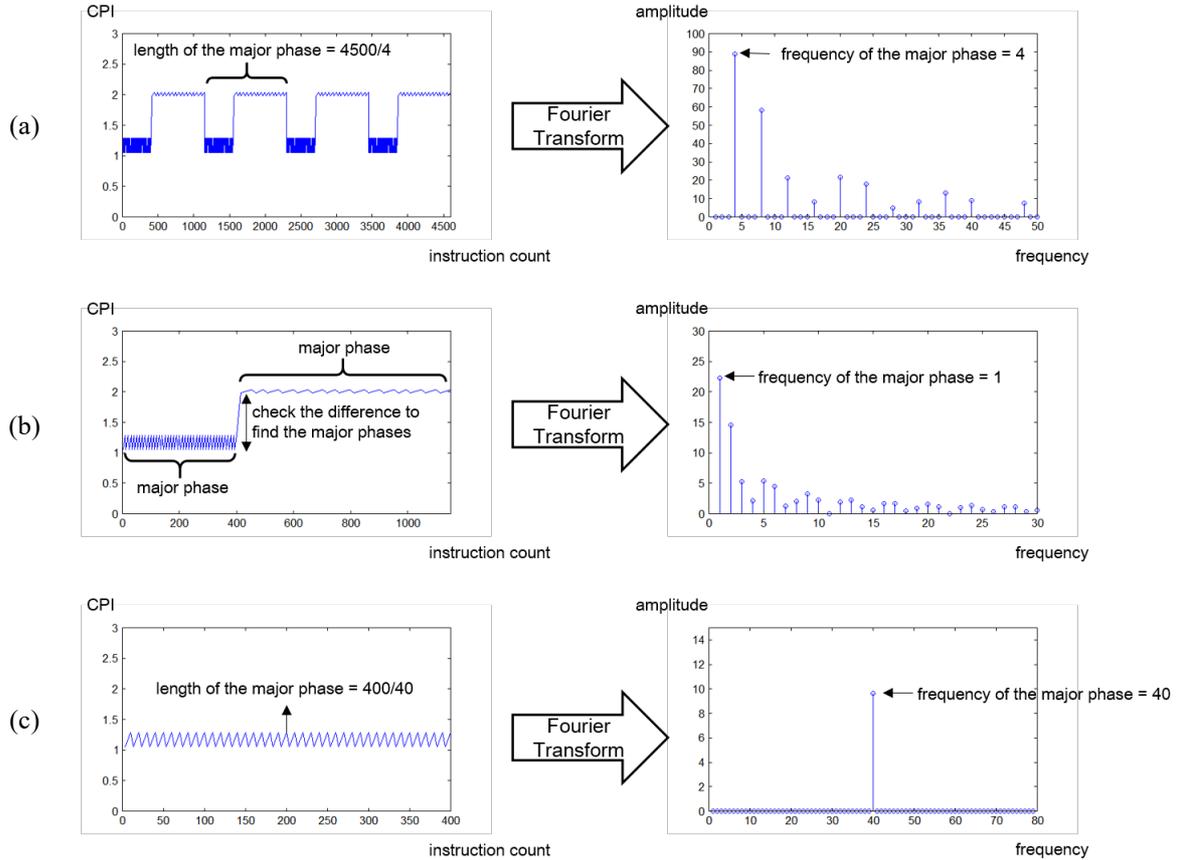

Figure 3. An example illustrates the proposed Frequency Domain Analysis method for systematic phase identification.

For the frequency domain spectrum shown in Figure 3, the horizontal axis indicates the pattern occurrence number in the time-domain execution period examined while the vertical axis represents the *dynamic* performance value. For clarity, we always ignore the first zero-occurrence value, which is the average performance value of the whole execution. In circuit analysis, this value is usually named the DC (direct current) value.

After removing the DC value, we locate the spectrum with the largest dynamic performance value, which is named the main spectrum and it shall correspond to the main program phase. We then use the occurrence value of the main spectrum to calculate the length of the main phase. For example, the occurrence value of the main spectrum in Figure 3(a) is 4, which means the main phase occurs four times. Since the total execution length is 4500 instructions, we have the main phase length equals to 1125 instructions, or 4500 instructions divided by 4. As illustrated by this example, the length of program phase can be identified much more easily in the frequency domain. We may formalize the phase length calculation as the following. Let $L$ be the main program phase length to be identified, $D$ the total length of the execution length and $X$ the occurrence number of the main spectrum. Then we have

$$L = D / X.$$

Using the length of the main phase as a hint, then we have in the next section how to combine with the code structure components (basic blocks, loops, functions) to identify precisely the phase starting point. Then the identified phase can be extracted from the time-domain waveform for further analysis.

As suggested by Fang [12], practically program phases are of a hierarchical structure. A primary phase may contain many secondary phases. Therefore, we simply apply a recursive approach to further analyze the primary phase identified using the same frequency domain transform approach and identify smaller phases within the primary phases. Recursively, we may find phases of all levels.

Note that for the special case as shown in Figure 3(b), when the main phase's occurrence is one, this case indicates no repeated pattern. Then we check if the difference between the performance measures of two neighboring phase segments are larger than the variance of the given waveform. Finally, we shall identify a high-phase and low phase boundary and have two separated phases.

Another special case occurs when the whole waveform is almost flat and the whole spectrum is also almost flat (after excluding the DC portion). This is the case that the whole

waveform is one single phase. Then we stop continuing recursive

As shown in Figure 3(c), after the third step recursive analysis, we may find a performance pattern which repeats every 10 (= 400/40) instruction count units, while the CPI variation is within 0.3. In this case, we may decide to treat this as a last level phase instead of exploring further.

Next, we discuss

*B. Bsic Block Performance Value*

To avoid the defects associated with the time-quantum approaches, we observe the fact that since the transition point separating any two program phases is always a branch instruction, a program phase can be fully decomposed into basic blocks. Although the quantum approach is still the easiest way to collect performance numbers, we develop a method that convert the measured performance numbers to each basic block.

Since there are no branch instructions inside a basic block, the number of instructions and instruction types are fixed. Therefore, the performance measure of each basic block is very consistent. To compute the performance value $p_b$ of a basic block $b$, we simply take the weighted average of the performance values of the time quanta that the basic block $b$ occurs in. Assume that the performance value of time quantum $q$ is $v_q$ and the basic block $b$ occurs $n_q$ times in the time quantum $q$. Then we have

$$p_b = (\Sigma_q n_q v_q)/(\Sigma_q n_q).$$

We have verified the so calculated basic block performance values and find them match very well with real values.

With the basic block performance values, we then redraw the performance waveform using the basic block instruction count for horizontal axis index instead of quantum sequence. With this block-block instruction indexed waveform, we essentially avoid the fuzzy phase boundary issue and can precisely set the phase boundary to align with a basic block. Nevertheless, the frequency domain analysis can still be applied with no modification to find phases.

Next, we discuss how to precisely identify phase starting point.

*C. Basic Block as Phase Starting Point*

Recall the fact that the starting instruction address of a basic block is always a target of certain branch instruction. Therefore, we may follow branch instructions during program execution to identify basic blocks at runtime efficiently instead of doing complicated control flow graph analysis. For convenience, we simply use the first instruction address of a basic block as its ID for easy identification.

With the unique basic block identification, then the phase detection problem is transformed into another problem of finding which basic block is the starting point of a phase. With the basic block performance values, we simply compare all consecutively occurring basic blocks and mark the pair whose difference exceed a given threshold, which is derived from the main phase's dynamic performance value. The second basic block of the pair with large difference is then taken as a candidate of the starting point of a phase.

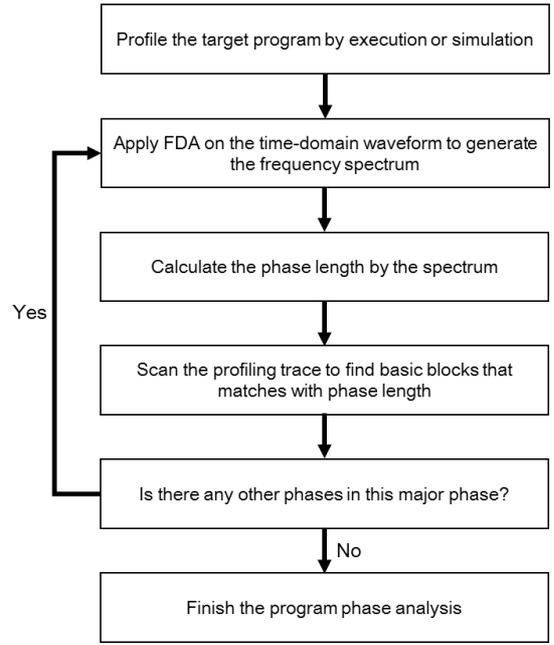

Figure 4. The flow diagram of the proposed approach.

Additionally, with the main phase length identified from the frequency domain analysis, we then check on the execution trace (time-domain waveform) with the hint of candidate basic blocks to precisely identify program phase.

*D. Code Structure and Phase*

In addition to having the advantage of effective and efficient identification of program phases using basic blocks, the basic block information can also be used to identify loops and functions. If knowing whether a phase is associated with loop or function code structure, system optimization algorithms can be more effective and can avoid unnecessary overhead since certain algorithms are more effective for loops and some are more for functions.

The starting point of a loop can easily be identified by checking whether the branch instruction is pointing backward, i.e. to a target instruction address which comes before the branch instruction address. Similarly, the starting point of a function can be identified by checking if the branch instruction is a function call instruction, such as the jump-and-link or JAL instruction in MIPS.

Since loops and functions are also identified through branch instructions, the starting points of loops and functions must be starting points of certain basic blocks. Therefore, we may annotate on each basic block whether it is also the starting point of a loop or function or none of them.

As discussed in previous sections, each identified program phase has a head basic block, which now also contains a code structure information. Therefore, we can easily find whether a phase encountered is related to loop or function.

*E. Hierarchical Phase Information*

By recursively executing the frequency domain analysis to identify phases, we then construct a hierarchical program phase table recording the phase's head basic block, length and performance value. Conversely, we also mark on each basic block which phase it is associated with.

Based on the above information, a program phase can be easily detected at runtime. Whenever a new basic block is executed, we first check if the basic block is the head of any program phase. If a phase is identified, we then look up the phase table and obtain the performance value (CPI), and code structure (loop or function) information.

Alternatively, with a customized instruction, the basic block phase information can also be embedded into the target application code. This special instruction can be inserted next to the first instruction of each head basic block. The instruction shall contain performance value of its associated phase and code structure information but function like an NOP instruction. Each new customized instruction encountered at runtime indicates the starting of a new phase.

*F. Propose Algorithm Flow*

Figure 4 summarizes the algorithm flow of our proposed method. First, we profile the target application either by actual execution or by simulation. At the same time, we obtain the basic block trace, CPI value of each basic block and the time domain performance waveform of the whole execution. Next, we apply the frequency domain analysis on the time-domain waveform and generate frequency spectrum. We take the main spectrum with the largest dynamic performance value and calculate the corresponding phase length. Then we scan through the time-domain waveform to find the head basic block of the main phase that matches with the phase length implied by the main spectrum. If the main spectrum occurrence number is one, then we identify the high-performance phase and the low performance phase. Only if the time-domain waveform is flat then we stop recursive search of the next level phase; otherwise, we recursively perform the frequency-domain analysis on the phase segment extracted from the time-domain waveform and continue to identify next level phase.

The proposed is implemented and verified with very encouraging results, which are discussed in the next section.

4. Experiments

*A. Methodology*

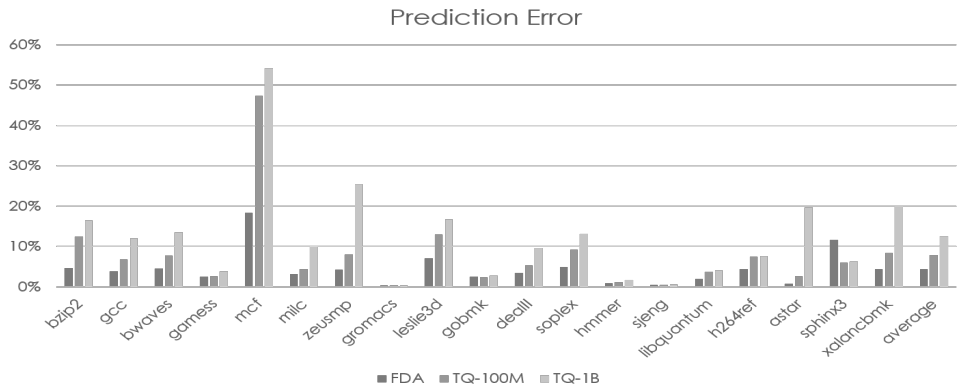

Figure 5. The error rate of program behavior prediction.

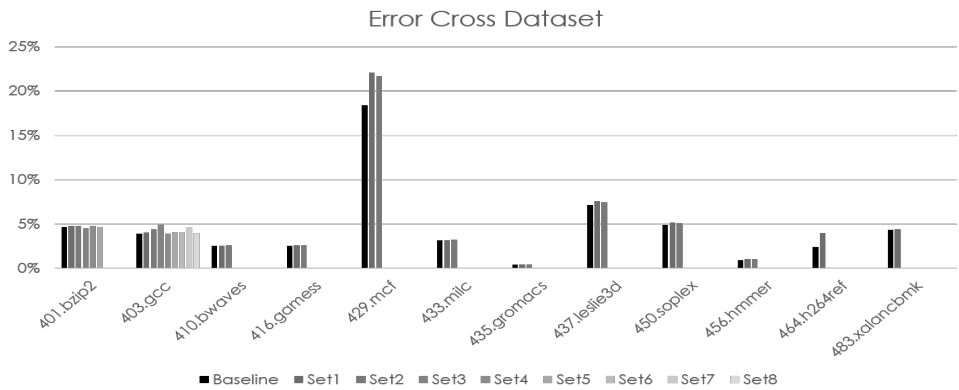

Figure 6. The error rate of different inputs in a program.

To verify our proposed prediction method, we adopt SPEC CPU2006 [17] benchmark suite as the target applications, and adopt the simulation model of SimpleScalar [16] to collect target program information. All experiments are performed on an *Intel Core-I5-3320m 3.30GHz* machine.

For the experiments, we modify SimpleScalar to collect basic blocks trace, compute CPI values of basic blocks for each test case. With the trace and CPI values, we apply the recursive frequency domain analysis to extract hierarchical program phases.

### B. Evaluation Results

In order to validate the accuracy of the proposed BBFDA approach, we also adopt the commonly used time-quantum length, 100 Million and 1 Billion instructions, and calculated the runtime CPI waveform for each test case for comparison. We then apply the proposed BBFDA method discussed in Section 3 using the CPI values annotated on the basic blocks to estimate the CPI waveform. Note that the basic block CPI values are computed onetime from a given reference input and the results are used for runtime performance estimation. For fair comparison, we use the golden simulation results as the baseline for error calculation.

Figure 5 displays the CPI estimation errors of our proposed approach (BBFDA) and the time-quantum based approaches (TQ-100M and TQ-1B). The results show that the average error rate of BBFDA is 4.45%, lower than that of TQ-100M (7.88%) and TQ-1B (12.54%). Therefore, the accuracy of the proposed BBFDA is roughly two to three times less than the time-quantum approaches and the BBFDA annotated phase and performance information can practically be used for system optimization purpose.

Specifically, the maximum error of BBFDA is 18.4% for mcf case and the least error is 0.42% for gromacs case. Mainly the difference is due to the variations of performance patterns. For example, for mcf case the CPI waveform exhibits large variations, hence the prediction is less accurate.

We also verify the effects of feeding different input data to the test cases and summarize the results in Figure 6. Since a reference input usually covers all functions of program, all program phases should be covered in our approach. Most benchmark cases show only slightly higher variations when taking different inputs. Therefore, with the basic-block-phase information, our approach is verified to be capable of identifying program phases under varying input data.

### C. Discussions

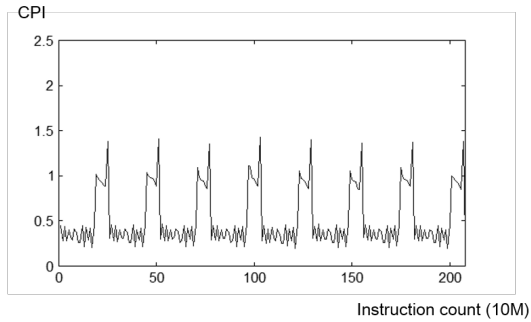

(a)

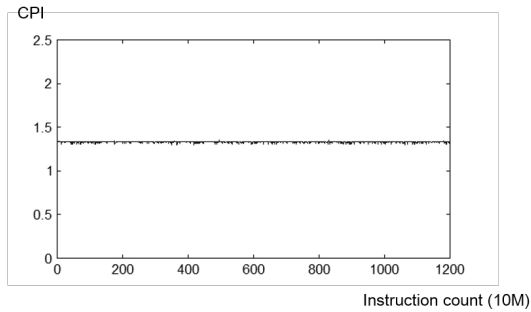

(b)

Figure 7. (a) The CPI waveform of the mcf benchmark case. (b). The CPI waveform of the gromacs benchmark case.

Shown in Figure 7(a) is the CPI waveform of the mcf benchmark case. Note that the lower CPI (0.2-0.5) phase exhibits large variations from the average value. Specifically, the high CPI value is approximately twice than the least value. The high variation case displays even more serious impact to the time-quantum based approach. For a larger sized time quantum, the CPI variation in general is higher since more components of different performance values are included for consideration. Therefore, the 1B quantum size shows higher error than the 100M quantum size. The average value of time quanta clearly cannot properly represent dynamic performance behaviors, whereas our proposed method can dynamically track the performance behaviors because we use the basic block analysis method and avoid the choice of granularity issue.

Figure 7(b) shows a case, gromacs, with little estimation error. Note that the case of interest is quite consistent throughout the execution. In this case, the time-quantum based method can obtain good results simply by providing an average value. Our basic-block based approach performs equally well in this case. Therefore, we may conclude that our approach is a robust approach.

We also observe that the performance estimation error rate is higher for those cases with many phases. The reason is that more phases implies more variations of performance behaviors. In general, our basic-block FDA approach performs more accurately than the traditional time-quantum based approach.

In Figure 8, we use the example to demonstrate that the proposed BBFDA method can precisely capture the length of program phases, so the BBFDA performance waveform shown in Figure 8(b) is similar to the simulation reference waveform shown in Figure 8(a). For the coarse time-quantum case of 260M in length shown in Figure 8(c), basically the waveform is flat and no phases can be identified since the quantum size is larger than all phase lengths. In contrast, for the time-quantum case of 80M in length shown in Figure 8(d), the phase boundaries are clearly distorted as compared to the golden reference. In conclusion, finding an appropriate quantum size is indeed a challenging job, if not impossible as we can see from the example.

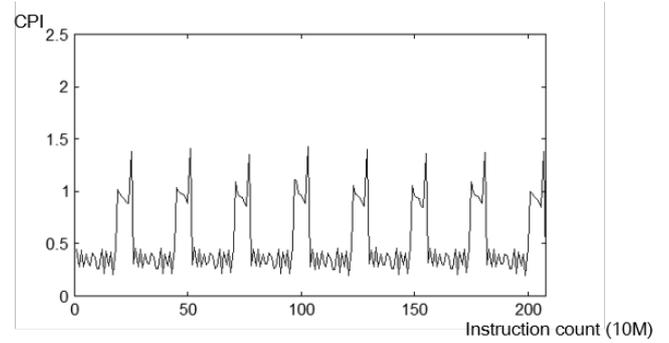

(a)

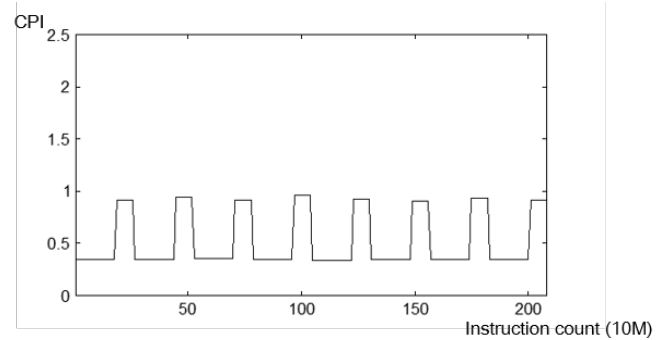

(b)

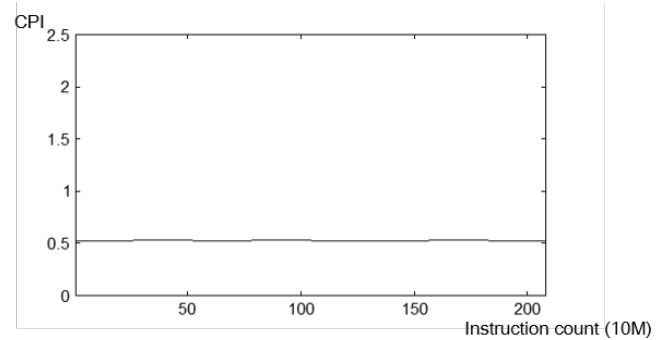

(c)

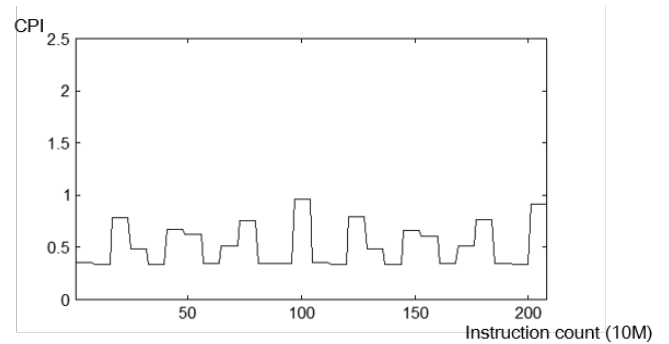

(d)

Figure 8. Comparing performance waveforms of the mcf case generated from various methods. (a) the fine-grained waveform; (b) the basic-block-based waveform; (c). the waveform of 260M-instruction time quantum; (d). the waveform of 80M-instruction time quantum.

## Conclusion

In this paper we present an efficient and effective frequency domain based analysis approach for program phase identification. The low frequency spectrum can be used to precisely find the major program phase. Our approach is more robust as we adopt the basic-block-based performance evaluation method and effectively avoid the granularity issue.

Experimental results show that our approach provides accurate performance estimations with an average of only 4.45% error rate as compared with the golden simulation references. Additionally our approach provides detailed hierarchical phase information annotated on basic blocks. For future work, the proposed method can be extended for facilitating runtime system optimizations.